\newcommand{\halpha}{H$\alpha$~}
\newcommand{\hbeta}{H$\beta$~}
\newcommand{\lyalpha}{Ly$\alpha$~}
\newcommand{\lymanb}{Ly$\beta$~}
\providecommand{\adsurl}[1]{\href{#1}{ADS}}%\documentclass[12pt,preprint]{aastex}}

\documentclass[iop,twocolumn]{emulateapj} 
\usepackage{graphics,graphicx,epsf,natbib}
\usepackage{subfigure}
\usepackage[colorlinks=true,linkcolor=blue,anchorcolor=blue,citecolor=blue,urlcolor=blue]{hyperref}
\usepackage{color}
\usepackage{graphicx}
\usepackage{amssymb,amsmath}

\usepackage{natbib}

\slugcomment{ApJ Accepted 4 Dec 2016}

\shorttitle{Signatures of Cooling and Star Formation}
\shortauthors{Donahue et al.}

\begin{document}


\title{Observations of \lyalpha and \ion{O}{6}: Signatures of Cooling and Star Formation in a Massive Central Cluster Galaxy}


\author{Megan Donahue\altaffilmark{1,2}, Thomas Connor\altaffilmark{1,3}
             G. Mark Voit\altaffilmark{1}, Marc Postman\altaffilmark{4} }
\altaffiltext{1}{Physics and Astronomy Dept., Michigan State University, East Lansing, MI, 48824 USA}
\altaffiltext{2}{donahue@pa.msu.edu }
\altaffiltext{3}{The Carnegie Observatories, 813 Santa Barbara Street, Pasadena, CA, 91101, USA}
\altaffiltext{4}{Space Telescope Science Institute, 3700 San Martin Drive, Baltimore, MD, 21218, USA}

\begin{abstract}
We report new HST COS and STIS spectroscopy of a star-forming region ($\sim100 M_\odot$ yr$^{-1}$) in the center of the X-ray cluster RXJ1532.9+3021 ($z=0.362$), to follow-up the CLASH team discovery of luminous UV filaments and knots in the central massive galaxy. We detect broad ($\sim 500$ km s$^{-1}$) \lyalpha emission lines with extraordinarily high equivalent width (EQW$\sim 200$\AA) and somewhat less broadened \halpha ($\sim 220$ km s$^{-1}$).  Emission lines of \ion{N}{5} and \ion{O}{6} are not detected, which constrains the rate at which  gas cools through temperatures of $10^6$~K to be $\lesssim 10$ M$_\odot$ yr$^{-1}$.   The COS spectra also show a flat rest-frame UV continuum with weak stellar photospheric features, consistent with  the presence of recently-formed hot stars forming at a rate of $\sim 10$ M$_\odot$ yr$^{-1}$, uncorrected for dust extinction. The slope and absorption lines in these UV spectra are similar to those of Lyman Break Galaxies at $z\approx 3$, albeit those with the highest \lyalpha equivalent widths and star-formation rates.  This high-EQW \lyalpha source is a high-metallicity galaxy rapidly forming stars in structures that look nothing like disks. This mode of star formation could significantly contribute to the spheroidal population of galaxies.   The constraint on the luminosity of any \ion{O}{6} line emission is stringent enough to rule out steady and simultaneous gas cooling and star formation, unlike similar systems in the Phoenix Cluster and Abell 1795. The fact that the current star formation rate differs from the local mass cooling rate  is consistent with recent simulations of episodic AGN feedback and star formation in a cluster atmosphere.

\end{abstract}

\keywords{galaxies: clusters: intracluster medium, galaxies:ultraviolet }

\section{Introduction}

A Brightest Cluster Galaxy (BCG) is typically extremely massive ($\sim10^{12}$ M$_\odot$) with a stellar
population dominated by old red stars, and lives in a cluster of galaxies permeated with hot ($10^7$~K), X-ray emitting 
intergalactic gas. Given their red-and-dead reputations, it might be rather surprising to learn that such galaxies can have
powerful emission-line regions and radio sources \citep{1985ApJS...59..447H,1999MNRAS.306..857C,1981ApJ...250L..59H}, cold dusty molecular gas 
\citep{2000ApJ...545..670D,2001MNRAS.328..762E}, and UV emission from
recent star formation ($\gtrsim 1-10 ~\rm{M}_\odot 
~\rm{yr}^{-1}$). Nature provides a strong clue as to why some of these galaxies are active and some are not: 
these phenomena, associated with active galactic nuclei (AGN) or star formation activity, only occur in clusters of galaxies
in which the hot, X-ray emitting gas within about 100 kpc of 
the center of the cluster has a cooling time short compared to $\sim 2$ Gyr, or equivalently, a central
gas entropy of $\lesssim 30$ keV cm$^2$ \citep{1989ApJ...338...48H,2008ApJ...683L.107C,Rafferty+08,2009ApJS..182...12C,2009ApJ...704.1586S}. 
There is mounting evidence that the condensation of hot gas fuels AGN and star-formation activity, a cycle
maintained by the considerable influence of the AGN on the intracluster medium (ICM). 

Recent rest-frame UV images obtained with the Hubble Space Telescope (HST) \citep{2015ApJ...805..177D,2015MNRAS.451.3768T} 
show that the ultraviolet emission in these active BCGs
has a lumpy and occasionally filamentary morphology, usually extending to around 10 kpc, but occasionally 
up to 100 kpc away from the center of 
the BCG. The emission-line morphologies of these galaxies in H$\alpha$, visible for nearby ($z<0.15$) BCGs in ground-based
images, is similar \citep{1985ApJS...59..447H}. 

The exact nature of these phenomena remains unclear. At least some of the gas is being photoionized by recent star
formation, which has been detected using diagnostics from the far-infrared through the UV. However, the full range
of emission-line ratios and linewidths cannot be easily explained by simple stellar photoionization \citep{1997ApJ...486..242V}. 
Astronomers have speculated about shocks in the molecular gas and particle heating associated with either the X-ray gas or 
relativistic particles associated with the radio sources \citep[e.g.,][]{2011MNRAS.417..172F}.  We and others have posited that 
the AGN lifts low-entropy but hot cluster gas out of the center of the galaxy, which then cools and rains back down towards
center \citep{2015Natur.519..203V, 2015ApJ...802..118B,2016ApJ...830...79M}. 

UV spectroscopy of moderately redshifted ($z\sim0.3-0.5$) sources 
affords the possibility of detecting 100,000 K gas in UV emission lines such as \ion{O}{6}~$\lambda \lambda$1032\AA,1038\AA, 
and \ion{C}{4}~$\lambda \lambda$1548\AA,1551\AA~ simultaneously with Ly$\alpha$. 
Large-aperture ($10\arcsec \times 20\arcsec$) spectroscopy from the {\em International Ultraviolet Explorer} 
showed strong \lyalpha in a dozen low-redshift ($z\lesssim0.1$) BCGs \citep{1992ApJ...391..608H}.  
Later, {\em Far Ultraviolet Spectroscopic Explorer} (FUSE) observations by \citet{2001ApJ...560..187O} and \citet{2006ApJ...642..746B}
 detected \ion{O}{6} in A2597 and A1795 respectively.
The FUSE apertures in these studies would have included any contributions from an AGN, so the \ion{O}{6} may be associated with AGN.
A very limited set of HST Cosmic Origins Spectrograph (COS) observations of non-AGN 
clumps in other galaxies have yielded detections of \ion{C}{4} and He II in 
M87 \citep{2009ApJ...704L..20S, 2012ApJ...750L...5S} and of \ion{O}{6} in a filament in Abell 1795 \citep{2014ApJ...791L..30M}. \ion{O}{6} 
 was detected from the center of the Phoenix cluster \citep{2015ApJ...811..111M}. 
In the M87 clump, no UV continuum sources were detected. M87 is close enough that 
{\em individual O-stars} would have been detected if they were present \citep{2009ApJ...704L..20S}.
In A1795, the \ion{O}{6} luminositiy corresponds to a mass of gas cooling radiatively at a rate $\sim  8$ times faster 
than the corresponding star-formation rate implied by the UV continuum. In the Phoenix cluster BCG, 
\citet{2015ApJ...811..111M} report that while some of the emission could come from shocks, gas could be cooling at
rates of over $1000 ~\rm{M}_\odot ~\rm{yr}^{-1}$ from a region where the star formation rate is $\sim700$~M$_\odot$~yr$^{-1}$.

To make simultaneous measurements of \ion{O}{6} and \ion{C}{4}  in a similar BCG with optical emission-line filaments and a relatively high rate of star formation,
we chose the BCG in the cluster RXJ1532.9+3021 at a redshift of $0.362$, which shifted the wavelengths of both of these emission lines to a
range that could be observed by COS.
We obtained COS FUV and NUV spectrosocpy, along with high-resolution UV imaging and long-slit spectroscopy with the Space Telescope Imaging Spectrograph (STIS).
We present these spectroscopic observations targeting the central region of the BCG. We also present a brief STIS long-slit observation centered
on the same region, covering the wavelength of \halpha at this redshift. 

Throughout the paper, we assume cosmological parameters of $\Omega_M=0.3$, $\Omega_\Lambda=0.7$, and $H_0=70$ km s$^{-1}$ Mpc$^{-1}$. 
At a redshift of 0.362, the scale is 5.0 kpc arcsec$^{-1}$ and the luminosity distance is 1920 Mpc \citep{2006PASP..118.1711W}.

\section{The cluster RXCJ1532.9+3021}

The central galaxy (RA(J2000), Dec(J2000) = 15:32:53.78,  +30:20:59.4) of the X-ray cluster RXCJ1532.9+3021 was first noted as a radio source (MGJ153246+3021) in the MIT-Green Bank II 5 GHz survey \citep{1990ApJS...72..621L}. A significantly more accurate radio location and flux density were measured by \citet{1995ApJ...450..559B}.   
The radio-loud active galaxy was associated with a ROSAT X-ray source by \citet{1997A&AS..122..235L}. The cluster itself was identified 
as a ROSAT X-ray source by \citet{1998MNRAS.301..881E} in the ROSAT Brightest Cluster Sample, and its spectroscopic 
redshift was first published by \citet{1999MNRAS.306..857C}. It was the target of Chandra observations \citep{2012MNRAS.421.1360H,2013ApJ...777..163H}.

The cluster was chosen to be a part of the Cluster Lensing And Supernovae survey with HST (CLASH) 
Multi-Cycle Treasury project as one of the 20 X-ray selected clusters.
The CLASH project \citep{2012ApJS..199...25P} collected HST Wide Field Camera 3 (WFC3) and Advanced Camera for Surveys (ACS) 
images in 16 bandpasses from the UV to the NIR 
for a sample of 25 massive clusters of galaxies with redshifts from 0.2-0.9. (The additional 5 clusters were included because they are
strong gravitational lenses of the high redshift univervse.) The
broad wavelength coverage insured sufficiently accurate photometric redshift estimates for lensed galaxies. 
As a result, the clusters themselves were imaged at UV wavelengths, observations usually denied to targets presumed to be red and dead. 
All of the CLASH BCGs were detected in the UV, some of them with UV colors indicating high star formation rates.
The UV analysis and images of the CLASH BCGs were published by \citet{2015ApJ...805..177D}.

\begin{figure*}
% this uses a figure trimmed by a tex script called:  pdfcrop --margins 10 fig.pdf fig.pdf
\includegraphics[width=\linewidth]{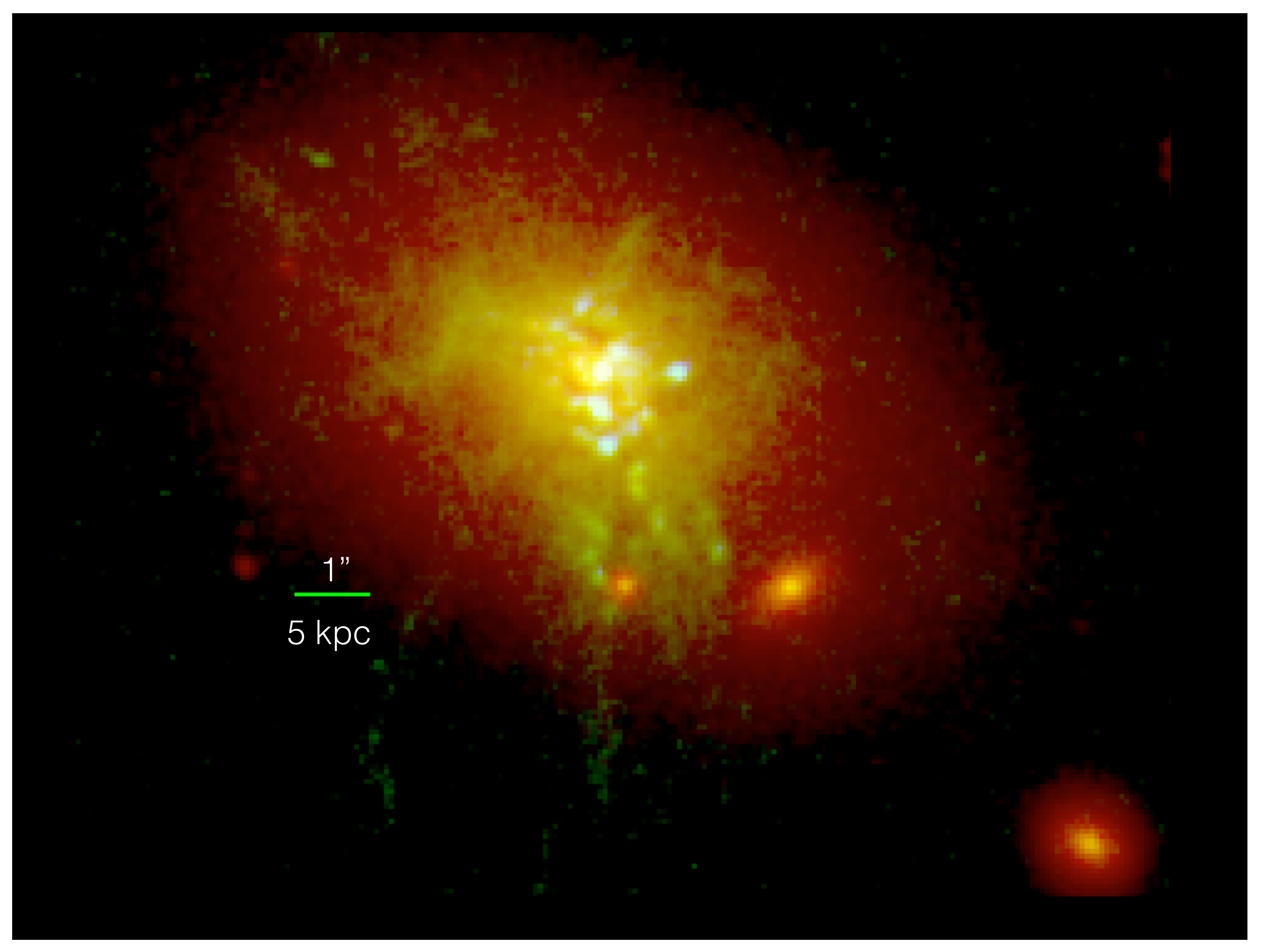}
\caption{\footnotesize
The RGB color image  shows the central $14\farcs4 \times 11\farcs2$ ($72 \times 56$ kpc) of the BCG 
 in RXJ1532.9+3021. North is up and East is to the left. The scale bar to the lower left is $1\arcsec$ (5 kpc). 
 The red color is from the HST WFC3/F140W image, dominated by stellar continuum emission. The green
  color is of the ACS/F475W image, a filter which includes the [\ion{O}{2}] emission line. 
The blue color is from the UV continuum image (WFC3/F275W) previously published in \citet{2015ApJ...805..177D}. 
\label{figure:HST}
  }
\end{figure*}

The BCG in RXJ1532.9+3021 (Shown in Figure~\ref{figure:HST}) 
bears a strong resemblance to other BCGs in strong cool cores with high star formation rates such as that in the Phoenix 
Cluster \citep{2012Natur.488..349M} or extended, filamentary optical nebular emission such as in NGC1275 \citep{1970ApJ...159L.151L}. It is located in the center of a luminous X-ray cluster with a highly peaked X-ray profile and relatively low central 
gas entropy. It has a significant amount of molecular gas. \citet{2012ApJS..199...23H} reports a detection with Spitzer FIR corresponding to a SFR similar
to that of a vigorous starburst ($>100$ solar masses per year), with an unobscured UV SFR that is nearly as high \citep{2015ApJ...805..177D}. A
detailed analysis of 2-d CLASH imaging and ground-based spectroscopy from \citet{2015ApJ...813..117F} estimates a total star formation rate of 
$120^{+220}_{-40}~\rm{M}_\odot~\rm{yr}^{-1}$. The \halpha filaments extend
beyond 50 kpc from the center of the BCG, and the UV appearance shows both knots and filaments. It has been postulated that the star formation
in these systems is fueled by gas condensing out of the ICM and is regulated by the kinetic energy deposited by
jets from the radio AGN \citep[e.g.,][]{2007ARA&A..45..117M}. However, local measurements of both cooling and star formation in BCGs have been
rare.

\begin{figure*}
% this uses a figure trimmed by a tex script called:  pdfcrop --margins 10 fig.pdf fig.pdf
\begin{center}
\includegraphics[width=6.0in]{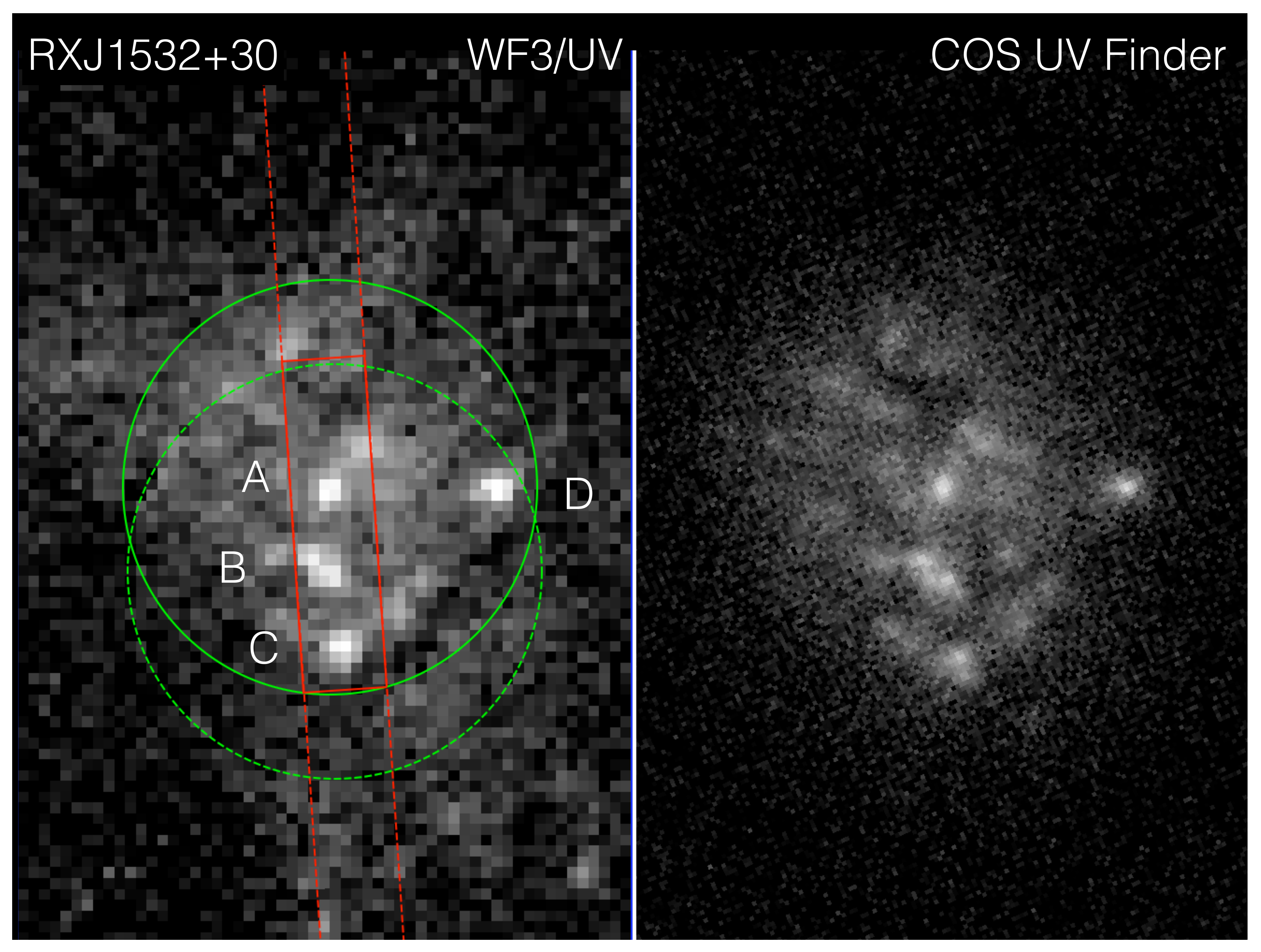}
\end{center}
\caption{\footnotesize
A grey-scale image of the central region of RXJ1532.9+3021 taken through the HST/WFC3 F275W filter is on the left and a stack of the higher-resolution COS UV acquisition images is on the right, with the same orientation and scale as the F275W image.  The solid green circle is the COS aperture (diameter $2\farcs5$). The dashed green circle shows the location of the COS aperture for the NUV observation. The red dashed rectangle shows the $0\farcs5$ wide STIS long-slit aperture. The solid red rectangle identifies the $2\farcs0$ range of the spectral extraction region for STIS, which was centered on the peak of \halpha emission (See Figure~\ref{figure:STIS2d}). 
 \label{figure:COSFinder}
  }
\end{figure*}

The CLASH UV images of the BCG in RXCJ1532.9+3021 showed distinct knots of emission. These knots are bright enough to be acquired and studied with
COS, which is optimized for the study of point-source spectra. Figure~\ref{figure:COSFinder} identifies each of these knots and the actual locations of the COS and STIS apertures. 
The HST proposal ID was 13367. The complete program included two STIS visits and three COS visits. The dates, exposure times, observation
strategies, and data reductions for each HST instrument are described in each sub-section of \S\ref{DATA}. We conclude that
section with a presentation of the extraordinarily high-resolution COS rest-frame UV acquisition images and a comparison to the CLASH photometry.
For the central region of this BCG, we measure the rest-frame UV luminosity, \lyalpha emission, \lymanb absorption, 
together with strong upper limits to \ion{O}{6} and weaker limits on \ion{C}{4} owing to its location on the 
vignetted end of the COS NUV spectral coverage. We estimate the unattenuated and total star formation rates based on the
UV continuum measurements and extinction estimates and \halpha emission line luminosities from STIS and the Sloan Digital Sky Survey (SDSS). 
We also estimate the rate at gas is cooling through a temperature of $10^6$~K from
our limits on \ion{O}{6} and conclude that the rate at which gas is cooling in the central region is less than the star formation rate in the same region. 
We discuss these results in \S\ref{DISCUSSION}, and summarize the paper in \S\ref{CONCLUSIONS}.

\section{Data Acquisition and Analysis \label{DATA}}

 In this section, we describe the observations, the preparation of the science products, and the spectroscopic measurements. 
To allow readers to skip the technical details, we have separated the measurements and the data preparation into separate 
but sequential subsections.

\subsection{STIS 2D Spectroscopy Observations and Data Reduction}

In order to obtain an \halpha  to \lyalpha ratio, we used STIS CCD/long slit
observations. We observed the central region through the $52\arcsec  \times 0.5\arcsec$ slit. 
The G750M grating was set  to a central wavelength of 8825~\rm{\AA}.
The first CCD STIS observation (ocap01), taken 18 March 2014, failed to acquire the target. The exposure time for the acquisition was too short, since, as a mildly resolved object, the target is extended. The initial HST pointing was quite accurate, but the acquisition routine moved the peak of the emission well off the STIS slit.
The second observation (ocap05), taken a little over one year later on 23 March 2015 successfully acquired the target. Two image sets were combined. The total exposure was 2,054 seconds. The position angle of the aperture was -176.22 degrees east of north (as shown in Figure~\ref{figure:COSFinder}).
 A fringe frame was obtained and used in calibration. The standard cosmic-ray rejection in the pipeline did a poor job of identifying and removing cosmic rays, which seriously affect STIS observations. Adjusting the rejection parameters to make multiple passes through the data with the STSDAS task {\em stis.ocrreject} only made marginal improvements in cosmic-ray rejection. Therefore, to clean the image, we used pixel replacement with the median of a box of $11 \times 11$ pixels. This procedure left behind a clearly detected but extended \halpha emission line and a very weakly detected [\ion{N}{2}] emission line (Compare the two sides of Figure~\ref{figure:STIS2d}).

For flux calibration, the median-filtered data were processed through the remaining  the pipeline steps for flux and dispersion calibration, using CALSTIS code version 2.40 (23-May-2012) and IRAF/Ureka version 1.4.4.3. The data were corrected for a  (small) heliocentric radial velocity (-12.5 km~s$^{-1}$). 

\subsection{STIS 2D Spectroscopy Meausurements}

The peak intensity of the extended \halpha emission was $\sim 5 \times 10^{-16}$ erg s$^{-1}$ cm$^{-2}$ $\rm{\AA}^{-1}$ arcsec$^{-2}$. The central wavelength of the emission line is consistent with \halpha at redshift $0.3613 \pm 0.0007$. 
The line emission is  velocity-broadened (500-600 km~s$^{-1}$  FWHM). It is clearly spatially resolved along the slit, but distinguishing individual clumps was not possible. There is weak suggestion of a blue-side asymmetry in the emission line, which would suggest possible outflow if it were real. 

\begin{figure*}
% this uses a figure trimmed by a tex script called:  pdfcrop --margins 10 fig.pdf fig.pdf
\begin{center}
\includegraphics[width=6.0in]{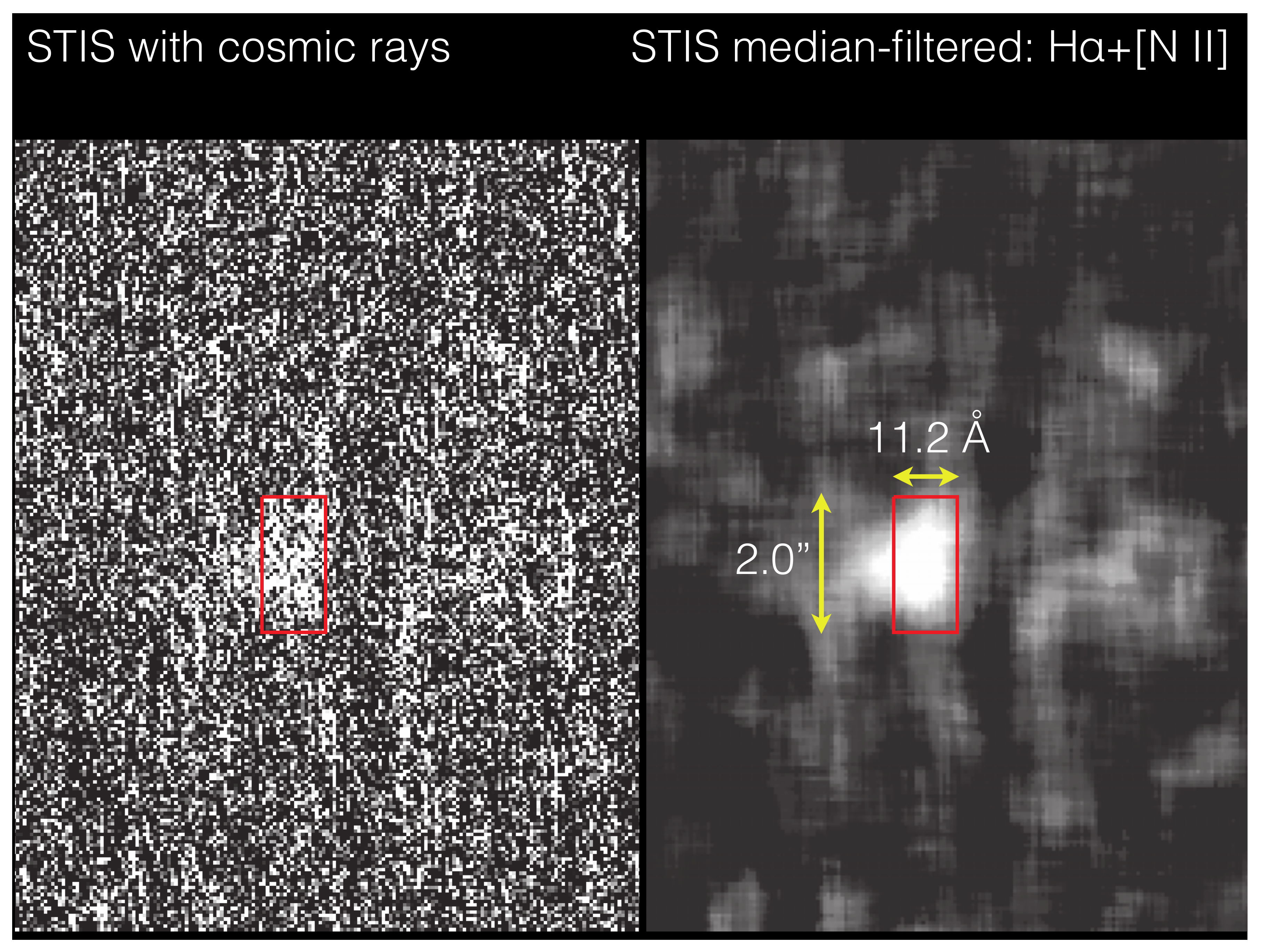}
\end{center}
\caption{\footnotesize
The grey scale images are the 2-D spectral data from STIS, plotting wavelength horizontally and location along the spectroscopic slit top to bottom.  On the left, we present a cosmic-ray-contaminated co-added image, and, on the right, the matching median-filtered image of the region near redshifted H$\alpha$.  Median filtering within boxes smaller than the scale of the emission-line feature effectively removed the cosmic ray contamination. The red boxes, identifying the region of the detector used to estimate the H$\alpha$ and [\ion{N}{2}] emission line features,} are identified on both images, corresponding to an approximate central wavelength of $\sim8936$ Angstroms and a width of 20 pixels ($11.2~\rm{\AA}$) and a height of 40 STIS CCD pixels ($\sim 2.0\arcsec$).  
\label{figure:STIS2d}
\end{figure*}

We employed two methods to measure the width and extent of the \halpha from the STIS data.  
Our first method used IRAF/{\it splot} to sum the spectra 40 pixels (2.0$\arcsec$) along the slit, similar to the box height  in the right hand side of Figure~\ref{figure:STIS2d}, and fit a single Gaussian line to the main emission line. This procedure yielded an estimated flux of $6-7 \times 10^{-15}$ erg s$^{-1}$ cm$^{-2}$, with a linewidth of 15-19 Angstroms FWHM. 
 In our second method, we wrote an IDL program, utilizing the {\em mpfit2dfun.pro} code of Craig Markwardt\footnote{\url{http://cow.physics.wisc.edu/~craigm/idl/idl.html}} to 
fit 2-D Gaussians to a subset of the 2-D flux-calibrated data centered around the redshifted \halpha+[\ion{N}{2}] complex. We fit the source in the spectral direction with a redshifted \halpha+[\ion{N}{2}] complex modeled by 3 Gaussian lines. We required the best-fit redshift and the Gaussian widths of the individual lines to be identical and  the flux of [\ion{N}{2}]6548 line to be 1/3 that of [\ion{N}{2}]6583. We used vacuum rest-wavelengths for \halpha (6564.614 \AA) and [\ion{N}{2}] (6585.27\AA~ and 6549.86\AA~  respectively.) We found that we recovered similar results whether median-filtering was included in the models or not. To confirm the uncertainties for our measurements, we generated 1000 random datasets from the original pipeline data together with the variance maps, median-filtered those images, and fit the random datasets in the same way as the original data. 
The best-fit \halpha flux was $(5.9 \pm 0.7)\times 10^{-15}$erg s$^{-1}$ cm$^{-2}$, with a  Gaussian width (sigma) of $0.62\pm 0.05 \arcsec$, corresponding to $3.1 \pm 0.25$ kpc at $z=0.36$. The linewidth FWHM is $15 \pm 1$ \AA ($\sim500$ km s$^{-1}$, or $\sigma  \sim210$ km s$^{-1}$). These values are similar to what we obtained from 
the simple {\it splot} estimates. The best-fit [\ion{N}{2}] flux is consistent with a flux of 1/3 H$\alpha$. The implied \halpha luminosity inside a $2.5 \times 10$ kpc$^2$ projected region 
is $2.6 \times 10^{42}$ erg s$^{-1}$, which makes this BCG one of the most luminous BCGs in \halpha. This luminosity corresponds to a star formation rate of
20 M$_\odot$ yr$^{-1}$ inside this limited region, using the conversion in \citep{1998ARA&A..36..189K}.

The extent and the separations between Knots A and B in the CLASH data indicate that emission from Knots A and
B are blended together in the STIS data. There is a hint of a shoulder of a source approximately 20\% fainter than the bulk of the emission along the slit axis. But the 2-D data are adequately fit by a single extended source, which includes both Knots A and B in its extent, so we conclude that Knots A and B are indistinct in \halpha emission.

\subsection{Sloan Digital Sky Survey Spectrum Comparison to STIS}

A Sloan Digital Sky Survey (SDSS) fiber spectrum, taken in 2003, is available for the larger region in the RXJ1532.9+3021 BCG,  in SDSS-Data Release 13 (SDSS-DR 13; \citet{2016arXiv160802013S}). (For reference, the ID is SDSS J153253.78+302059.4.) The \halpha flux in the fiber is $2.5\pm0.2 \times 10^{-14}$, and the [\ion{N}{2}]6584/\halpha ratio is $0.74\pm0.07$ \citep{2013MNRAS.431.1383T}. Therefore, our $0.5\arcsec \times 2.0\arcsec$ extraction from the full 
STIS long-slit spectrum represents about 1/4 of the optical emission-line light detected by SDSS within $r=1.5\arcsec$ of the center of the galaxy. 
(We note the SDSS fiber flux is uncertain since a previous extraction, using the MPA-JHU pipeline, used in DR8 \citep{2011ApJS..193...29A}, estimates a flux in the \halpha line for this target that is about two times lower than that cited above. The [\ion{N}{2}]/\halpha ratios are consistent, however.)
Both SDSS analyses found linewidths via joint fits of many emission lines of 
$\sigma \sim 200$ km~s$^{-1}$, while the linewidth for \halpha from the knots is $\sigma_{STIS} \sim 220$ km~s$^{-1}$  (corresponding to a FWHM of $\sim500$ km~s$^{-1}$.) The higher [\ion{N}{2}]/\halpha ratio in the fiber spectrum suggests that the 
extended light may have a higher [\ion{N}{2}]/\halpha ratio ($\sim 1.2$) than the knots.
In summary, the SDDSS analysis of the existing fiber ($d=3\farcs0$) spectrum for RXJ1532.9+3021 and our analyses of the central $0\farcs5 \times 2\farcs0$ spectrum from STIS find similar redshift and line widths, but the SDSS spectrum shows somewhat larger [\ion{N}{2}]/\halpha ratios ($\sim0.8$).
 
\subsection{COS Spectroscopic Observations and Data Reduction}

To assess the presence and fluxes of FUV and NUV emission lines from ionized and hot gas in the BCG of RXJ1532.9+3021, we targeted Knot A (Figure~\ref{figure:COSFinder}) 
with the HST COS spectrograph. 
The acquisition and observation of a compact but extended source with a faint UV continuum and prominent emission lines is difficult with COS. The COS spectrograph has a $2\farcs5$ diameter aperture, but it is optimized for a compact source located in the center of the aperture. Prior to proposing for COS observations, we had acquired HST UV images from ACS with very precise knot locations (Donahue et al. 2015). 
But the pointing of HST is limited by the accuracy of whatever guide stars happen to be used for a given observation and roll angle. 

The first two acquisitions, in June of 2014, successfully centered on Knot A. We chose conservative exposure times of 800-1000 seconds for the acquisitions and a slight offset to the north and east of Knot A to ensure that the usual peak identification would choose Knot A over the other, less UV-bright knots. However, in December of 2014, the roll-angle was reversed, and the acquisition routine, although identical to the previous acquisitions (1000 s, same offset strategy), chose Knot B over Knot A. 
The same dominant Guide Star was used for all 3 acquisitions, but the December observation used a different Guide Star to control for its roll angle. We verified this difference by inspecting the header of the {\em .jif} files.
We report results for all spectra. Because knots A and B are separated by only $0\farcs5$, all COS spectra for this program include UV flux from both knots A and B and the surrounding region based on comparisons of the detected flux to CCD photometry from the ACS. The transmission through the COS aperture drops with offset from the center, but 
the transmission is higher than 90\% for the central $r\lesssim 0.5\arcsec$, and is still above 80\% for the central $r\lesssim 0.7\arcsec$. 
We show in \S~\ref{section:WFC3} by comparison to WFC3 UV photometry that the fluxes in the COS spectra are consistent with coming from the central $\sim2\arcsec$ diameter region in the core of the galaxy and are not dominated by flux from any individual knot, at least in the NUV where a direct comparison can be made.

A standard spectral extraction for the COS data was sufficient for the scientific needs of the project. An optimized FUV extraction using the IDL COS coaddition routine 
from the University of Colorado\footnote{\url{https://cos.colorado.edu/COS_idt.html}} 
resulted in a very similar spectrum. The calibration of COS spectra is accurate to $\sim 5\%$, with relative flux accuracies of about $2\%$ (COS Data Handbook Cycle 24.)

\subsection{COS Spectroscopic Measurements}

All of the COS observations detected faint UV continuum. The flux at observed 1500\AA~ is $F_{1500} \sim 9\times10^{-16}$~erg~s$^{-1}$~cm$^{-2}$~s$^{-1}$. This UV rate corresponds to a SFR of 3-10 solar masses per year, uncorrected for dust (the range presented 
here emerges from a range assumptions of the initial mass function of stars and the timescales, see 
\citealt{2013seg..book..419C}.) 
Only the FUV observation revealed any line emission, and only \lyalpha was detected. 
Neither \ion{O}{6} or \ion{C}{4} were detected in emission (although the redshift of \ion{C}{4} line put it very close to the edge of the vignetted, noisier 
limit of the NUV wavelength coverage.)

\begin{figure*}
% this uses a figure trimmed by a tex script called:  pdfcrop --margins 10 fig.pdf fig.pdf
\begin{center}
\includegraphics[width=6.0in]{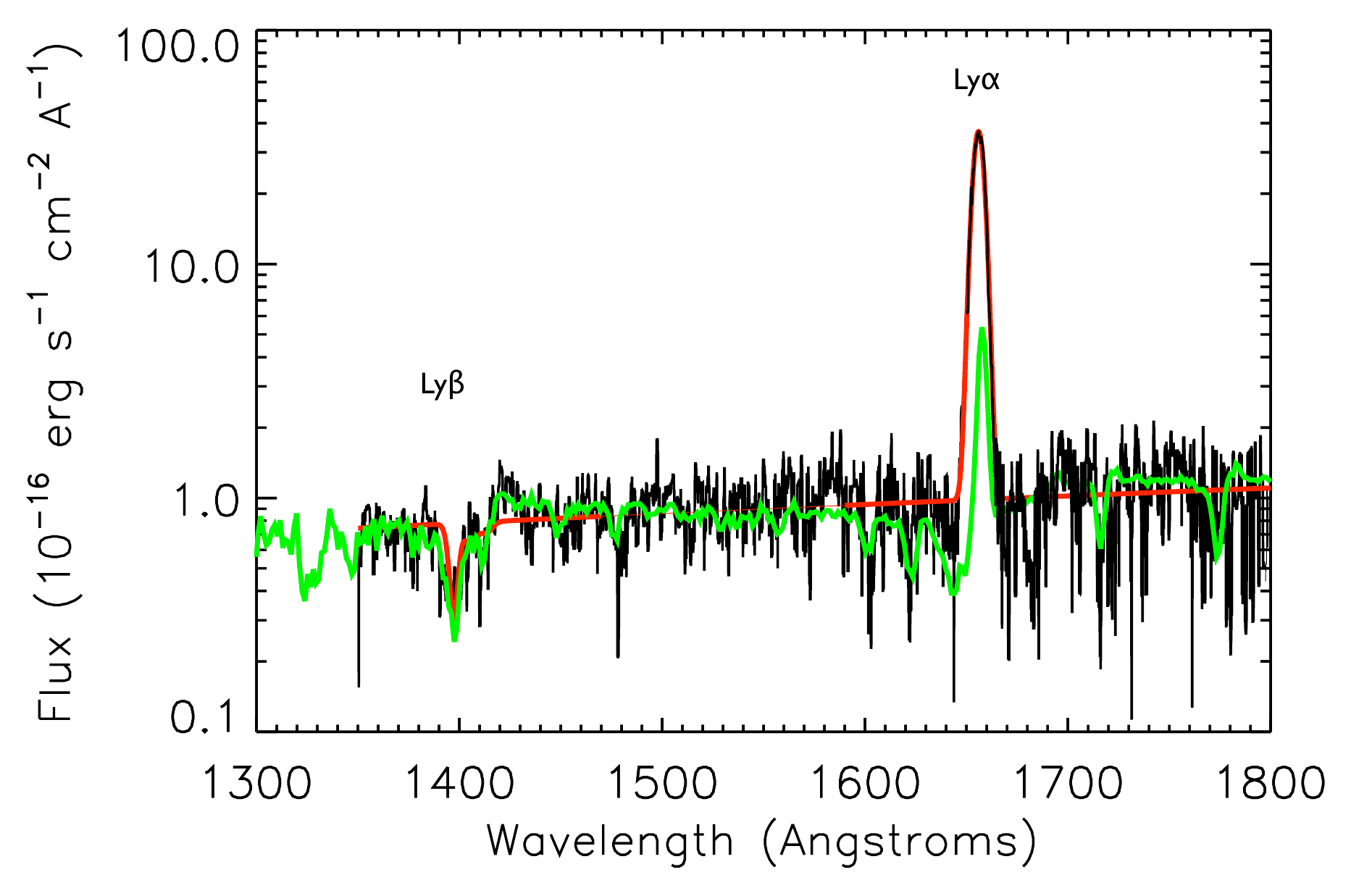}
\end{center}
\caption{\footnotesize
COS FUV spectrum, median filtered every 11 bins (0.88 \AA). The red line is the best-fit spectrum with Ly~$\alpha$ emission,
\lymanb absorption,  \ion{O}{6} (the best fit calls for absorption) and a power law continuum, normalized at 1600 \AA. 
The green line shows a mean template spectrum based on rest-UV observations of several hundred $z\sim3$ Lyman Break Galaxies from \citet{2003ApJ...588...65S}. 
The strong emission line at \lyalpha, the absorption line at \lymanb have been labelled. Stellar photospheric lines, such as \ion{C}{3} $\lambda$1176 (here seen near 1600\AA), 
are visible in alignment with the stacked LBG spectrum. 
 \label{figure:COS}
  }
\end{figure*}

\lyalpha was detected at a redshift of $0.36218$ (central observed wavelength of $1655.96\pm0.06\rm{\AA}$ 
with a flux of $(2.6 \pm 0.13) \times 10^{-14}$ erg s$^{-1}$ cm$^{-2}$, corresponding to an rest-frame equivalent width of 210\AA~ 
(Figure~\ref{figure:COS}). The line is 
broadened, $\sim 2.9$ \AA~ observed wavelength, corresponding to  $\sigma \sim 520 \pm 10$ km~s$^{-1}$ . 
\lymanb was detected in absorption at a nearly $4-\sigma$ level. When constrained to have the same redshift and line width as \lyalpha, the
amplitude of the absorption line was $(4.6 \pm 1.3) \times 10^{-17}$ erg s$^{-1}$ cm$^{-2}$ \AA$^{-1}$.
This amplitude corresponds to an equivalent width of $3.6 \pm 1.0 ~\rm{\AA}$. 
% EXPLICIT CALCULATION RETAINED IN COMMENTS -NOTES 
% For \lymanb, the computed oscillator strength is 0.079 based on an atomic
% transition probability of $5.578 \times 10^8 $ s$^{-1}$ and upper level statistical weight of 18, lower level statistical weight of 2. In the 
% optically-thin linear regime, the relationship between equivalent width and column density is $W_\lambda = 8.85 \times 10^{-5} \lambda(\rm{\AA})^2 f_{lu} N_{l}$, where
% the rest wavelength of the transition $\lambda$ and equivalent width $W_\lambda$ are in Angstroms, $f_{lu}$ is the oscillator strength (dimensionless), and $N_{l}$ is
%the column density in cm$^{-2}$ \citep{1968dms..book.....S}. 
The corresponding column density of neutral hydrogen is $4.9 \times 10^{15}$ cm$^{-2}$, or 120 solar masses of neutral hydrogen per square kpc.

The ratio of \lyalpha from the COS observation to \halpha measured with STIS is $\sim 4$, significantly less than what 
one would expect from recombination without correcting for intrinsic extinction. 
However, the extinction inferred from this ratio is moderate, corresponding to an $E(B-V)\sim 0.1-0.2$ for
a wide range of expectations for the intrinsic \lyalpha$/$\hbeta ratio.\footnote{Assuming $R=3.1$ here. 
The intrinsic ratio of \lyalpha to \halpha of 8.3 predicted from Case B recombination implies the least extinction; alternative 
intrinsic ratios of up to 25 yield the upper limit to extinction estimates.} This extinction is consistent with the
total reddening to the source estimated from the SDSS spectrum  ($E(B-V) =0.19\pm0.02$ mag) 
and is somewhat lower than that obtained through the analysis of CLASH photometry 
of a much larger region by \citet{2015ApJ...813..117F} ($E(B-V)\sim 0.25\pm0.04$, depending on exact assumptions). The Galactic
contribution to extinction towards this high Galactic latitude source is $E(B-V)=0.026$, based on maps and computations in 
\citet{2011ApJ...737..103S}, so all of these estimates demonstrate that a significant fraction of the total extinction in these
spectra is intrinsic to the BCG. 

The \lyalpha line (1215.67\AA) is well-fit by a simple Gaussian profile, and so there is no statistically significant evidence for asymmetry ($\chi_{red}^2 \sim 1$). 
The line width implies a broader velocity profile than that represented by 
the optical lines measured with STIS and SDSS. It is possible that
extended emission in the aperture broadens the effective line. However, the most likely explanation of
the broader line width is that the 
width of the \lyalpha line is enhanced by scattering from fast-moving neutral H atoms. Scattering 
affects \lyalpha much more than H$\alpha$. 
The nominal \halpha best-fit redshift, from the STIS observations ($z=0.3612 \pm 0.0007$),  is marginally closer to that of the stars as measured from the SDSS spectrum  ($z_{stars}=0.3619$), 
than that of the \lyalpha line ($z_{\rm{Ly}\alpha}=0.36218\pm0.00005$). The difference in mean redshift
and the larger difference in linewidths may arise if the \lyalpha light is scattered by a medium
that is transparent to  \halpha photons. 

The \ion{O}{6} emission line doublet (vacuum rest wavelengths 1031.97\AA, 1037.63\AA) 
was not detected in the COS/FUV spectrum, and the quality of the data allowed us to 
place a stringent upper limit on it. 
The three-sigma upper limits for \ion{O}{6}1033/\lyalpha $< 0.009$ correspond to a limit on gas cooling and 
condensation  from a hot phase of $<7-8$ solar masses per year \citep{1994ApJS...95...87V}.

There is a depression with a marginal statistical significance ($2-\sigma$) near \ion{C}{4} (1548.20\AA,1550.77\AA)
on the short wavelength edge of 
the NUV spectrum of the region. 
If the continuum from the region is primarily from recently-formed hot stars, some absorption is expected, although probably not at
the level that could be detected in our observations. 
A  $3-\sigma$ upper limit to emission at redshifted \ion{C}{4} is $<1.7 \times 10^{-16}$ erg s$^{-1}$ cm$^{-2}$, 
over 20 times fainter than the $4 \times 10^{-15}$ erg s$^{-1}$ cm$^{-2}$ that would be 
expected if Knot B had line ratios relative to \halpha similar to the off-nuclear emission-line knot studied in M87 \citep{2012ApJ...750L...5S}.

\ion{N}{5} (1238.8\AA, 1242.8\AA) was not detected, with a 3-sigma limit of $7.2 \times 10^{-16}$ erg s$^{-1}$ cm$^{-2}$. Since the ratio of \ion{O}{6}/\ion{N}{5} is about 10 in condensing gas, our \ion{N}{5} limit on condensing gas is 10 times weaker than our limit from \ion{O}{6}. The lack of \ion{N}{5}, \ion{O}{6}, and \ion{C}{4} in emission is also consistent with the lack of an optically luminous AGN.

\subsection{WFC3 and COS Finder Photometry \label{section:WFC3}}

The broadband UV photometry we present here shows that the continuum levels in the COS aperture exceed the light
emitted by the individual knots.  We also discuss here the source structures revealed by the co-addition of the
individual COS acquisition images.  The COS acquisition camera has the best sampling of the HST UV PSF, so 
its broad-band UV imaging represent some of the highest resolution UV images ever taken. (However, its
photometric usefulness is limited, for reasons we also discuss.).

We identified four knots in CLASH WFC3/UVIS images (Figure~\ref{figure:COSFinder}). We label them Knots
A, B, and C starting from the center of the BCG and
moving south; Knot D is located to the west. All 16 CLASH images are measured using the IRAF task {\it aphot}.
Additionally, we measured the fluxes for two large apertures that included all four knots, centered
on Knots A and B. The coordinates of the four knots, as well as
the center of the larger regions, the radii of the apertures used for photometry, and the
sizes of the background annuli used for local background subtraction are given in Table~\ref{table:Clash_Obs_param}. 

To measure the photometry of each knot, we used IRAF/{\it apphot}. The flux inside an aperture was found via
\begin{equation}
F = (\Sigma - N_A  B_{sky})/t_{exp},
\end{equation}
where $\Sigma$ is the total counts measured inside of the aperture, $N_A$ is the area of that aperture in pixels,  $B_{sky}$ is the measured background in counts per pixel, and  $t_{exp}$ is the exposure time. 
%Here, the background is measured using the median statistic in a $3\arcsec$-wide annulus centered on Knot A and starting $3\arcsec$ out. 
For a background region of $N_{sky}$ pixels, the error in the flux measurement in units of net counts per second is:
\begin{equation}
\sigma_F = \frac{1}{t_{exp}} \sqrt{ Ft_{exp} + N_A \sigma_{sky}^2 + \frac{N_A^2 \sigma_{sky}^2}{ N_{sky} } },
\end{equation}
where $\sigma^2_{sky}$ is the measured variance in units of counts$^2$ per area inside the background region.  We corrected the measured fluxes and error in the fluxes for Galactic extinction, and converted our measurements to AB magnitudes. We report our measured photometry for each filter and knot combination, as well as the corresponding filter zeropoint, linear flux correction applied for Galactic extinction, and exposure time $t_{exp}$ for each filter, in Table \ref{table:Phot}.

%The reported photometry uses a global background, not a local one
%\begin{deluxetable*}{lcccc}
%\tablecaption{CLASH Knot Photometry Parameters \label{table:Clash_Obs_param}}
%\tablehead{ \colhead{Knot Designation} & \colhead{$\alpha_{2000}$} & \colhead{$\delta_{2000}$} & \colhead{Aperture radius} & \colhead{Background Radii} \\
%\colhead{} & \colhead{} & \colhead{} & \colhead{($\arcsec$)} & \colhead{($\arcsec$)}  }
%A   & 15:32:53.772 & +30:20:59.43 & 0.3 & 0.3 - 0.6 \\
%B   & 15:32:53.775 & +30:20:58.94 & 0.3 & 0.3 - 0.6  \\
%C   & 15:32:53.765 & +30:20:58.49 & 0.3 & 0.3 - 0.6  \\
%D   & 15:32:53.695 & +30:20:59.40 & 0.3 & 0.3 - 0.6  \\
%All-A & 15:32:53.772 & +30:20:59.43 & 1.0 & 1.0 - 3.0 \\ 
%All-B & 15:32:53.775 & +30:20:58.94 & 1.0 & 1.0 - 3.0  \\
%\enddata
%\end{deluxetable*}

\begin{deluxetable}{lccc}
\tablecaption{CLASH Aperture Parameters \label{table:Clash_Obs_param}}
\tablehead{ \colhead{Knot} & \colhead{$\alpha_{2000}$} & \colhead{$\delta_{2000}$} & \colhead{Radius}  \\
\colhead{ID} & \colhead{} & \colhead{} & \colhead{($\arcsec$)}  }

A   & 15:32:53.772 & +30:20:59.43 & 0.3  \\
B   & 15:32:53.775 & +30:20:58.94 & 0.3   \\
C   & 15:32:53.765 & +30:20:58.49 & 0.3   \\
D   & 15:32:53.695 & +30:20:59.40 & 0.3   \\
All-A & 15:32:53.772 & +30:20:59.43 & 1.0 \\ 
All-B & 15:32:53.775 & +30:20:58.94 & 1.0   
\enddata
\end{deluxetable}

\begin{deluxetable*}{lrrrrrrrrr}
\tablecaption{Broadband Aperture Photometry \label{table:Phot}}
\tablehead{
\colhead{Filter} & \colhead{Knot A} & \colhead{Knot B} & \colhead{Knot C} & \colhead{Knot D}  & \colhead{Large A} & \colhead{Large B}  & \colhead{Zeropoint} & \colhead{$f$} & \colhead{$t_{exp}$} \\
\colhead{} & \colhead{(mag)} & \colhead{(mag)} & \colhead{(mag)} &  \colhead{(mag)} &  \colhead{(mag)} &  \colhead{(mag)} & \colhead{(mag)} & \colhead{} & \colhead{(s)}
}
\startdata
F225W  & $ 23.250 \pm 0.021 $  & $ 23.300 \pm 0.022 $  & $ 23.442 \pm 0.024 $  & $ 23.525 \pm 0.026 $  & $ 21.075 \pm 0.009 $  & $ 21.229 \pm 0.010 $  & 24.097 & 1.225 & 7262.00 \\ 
F275W  & $ 23.079 \pm 0.016 $  & $ 23.189 \pm 0.018 $  & $ 23.364 \pm 0.021 $  & $ 23.405 \pm 0.021 $  & $ 20.950 \pm 0.007 $  & $ 21.113 \pm 0.008 $  & 24.174 & 1.182 & 7428.00 \\ 
F336W  & $ 22.664 \pm 0.010 $  & $ 22.830 \pm 0.012 $  & $ 23.148 \pm 0.015 $  & $ 23.257 \pm 0.016 $  & $ 20.659 \pm 0.005 $  & $ 20.781 \pm 0.006 $  & 24.645 & 1.148 & 4764.00 \\ 
F390W  & $ 22.437 \pm 0.006 $  & $ 22.691 \pm 0.007 $  & $ 23.072 \pm 0.009 $  & $ 23.383 \pm 0.011 $  & $ 20.513 \pm 0.003 $  & $ 20.615 \pm 0.003 $  & 25.371 & 1.131 & 4840.00 \\ 
F435W  & $ 22.246 \pm 0.006 $  & $ 22.553 \pm 0.008 $  & $ 23.018 \pm 0.011 $  & $ 23.283 \pm 0.014 $  & $ 20.370 \pm 0.003 $  & $ 20.462 \pm 0.004 $  & 25.666 & 1.118 & 4100.00 \\ 
F475W  & $ 21.862 \pm 0.003 $  & $ 22.192 \pm 0.004 $  & $ 22.653 \pm 0.006 $  & $ 22.987 \pm 0.008 $  & $ 19.982 \pm 0.002 $  & $ 20.061 \pm 0.002 $  & 26.059 & 1.107 & 4128.00 \\ 
F606W  & $ 21.211 \pm 0.002 $  & $ 21.709 \pm 0.003 $  & $ 22.306 \pm 0.004 $  & $ 22.606 \pm 0.005 $  & $ 19.350 \pm 0.001 $  & $ 19.449 \pm 0.001 $  & 26.491 & 1.083 & 4060.00 \\ 
F625W  & $ 21.056 \pm 0.002 $  & $ 21.598 \pm 0.004 $  & $ 22.211 \pm 0.006 $  & $ 22.518 \pm 0.007 $  & $ 19.211 \pm 0.001 $  & $ 19.315 \pm 0.001 $  & 25.907 & 1.075 & 4128.00 \\ 
F775W  & $ 20.771 \pm 0.002 $  & $ 21.365 \pm 0.004 $  & $ 22.076 \pm 0.006 $  & $ 22.370 \pm 0.008 $  & $ 18.917 \pm 0.001 $  & $ 19.028 \pm 0.001 $  & 25.665 & 1.056 & 4090.00 \\ 
F814W  & $ 20.488 \pm 0.001 $  & $ 20.998 \pm 0.002 $  & $ 21.670 \pm 0.003 $  & $ 22.002 \pm 0.004 $  & $ 18.640 \pm 0.001 $  & $ 18.727 \pm 0.001 $  & 25.943 & 1.051 & 8036.00 \\ 
F850LP & $ 20.108 \pm 0.002 $  & $ 20.533 \pm 0.002 $  & $ 21.174 \pm 0.004 $  & $ 21.507 \pm 0.005 $  & $ 18.254 \pm 0.001 $  & $ 18.322 \pm 0.001 $  & 24.842 & 1.041 & 8134.00 \\ 
F105W  & $ 20.262 \pm 0.002 $  & $ 20.793 \pm 0.002 $  & $ 21.549 \pm 0.004 $  & $ 21.880 \pm 0.005 $  & $ 18.392 \pm 0.001 $  & $ 18.486 \pm 0.001 $  & 26.269 & 1.028 & 2814.67 \\ 
F110W  & $ 20.105 \pm 0.001 $  & $ 20.661 \pm 0.002 $  & $ 21.429 \pm 0.003 $  & $ 21.732 \pm 0.004 $  & $ 18.252 \pm 0.001 $  & $ 18.349 \pm 0.001 $  & 26.825 & 1.024 & 2514.67 \\ 
F125W  & $ 19.998 \pm 0.001 $  & $ 20.572 \pm 0.002 $  & $ 21.360 \pm 0.003 $  & $ 21.660 \pm 0.004 $  & $ 18.164 \pm 0.001 $  & $ 18.260 \pm 0.001 $  & 26.247 & 1.021 & 2514.67 \\ 
F140W  & $ 19.798 \pm 0.001 $  & $ 20.365 \pm 0.002 $  & $ 21.166 \pm 0.003 $  & $ 21.497 \pm 0.004 $  & $ 17.983 \pm 0.001 $  & $ 18.072 \pm 0.001 $  & 26.452 & 1.017 & 2311.74 \\ 
F160W  & $ 19.686 \pm 0.001 $  & $ 20.234 \pm 0.001 $  & $ 21.052 \pm 0.003 $  & $ 21.411 \pm 0.003 $  & $ 17.864 \pm 0.001 $  & $ 17.949 \pm 0.001 $  & 25.956 & 1.013 & 5029.34 \\ 
\enddata
\tablecomments{ Total AB magnitudes based on HST imaging inside apertures defined in Table~\ref{table:Clash_Obs_param}. A constant global background was subtracted, but local BCG light is included. The corresponding zeropoint assumed for each filter, linear flux correction $f$ for Galactic attenuation ($F_{corrected}= f F$, $F$ as defined in Equation 1), and total exposure time ($t_{exp}$) are reported for completeness.}

%% General table references marker
%\tablerefs{\citet{2012ApJS..199...25P}}

\end{deluxetable*}

\begin{figure*}
% this uses a figure trimmed by a tex script called:  pdfcrop --margins 10 fig.pdf fig.pdf
\includegraphics[width=6.0in]{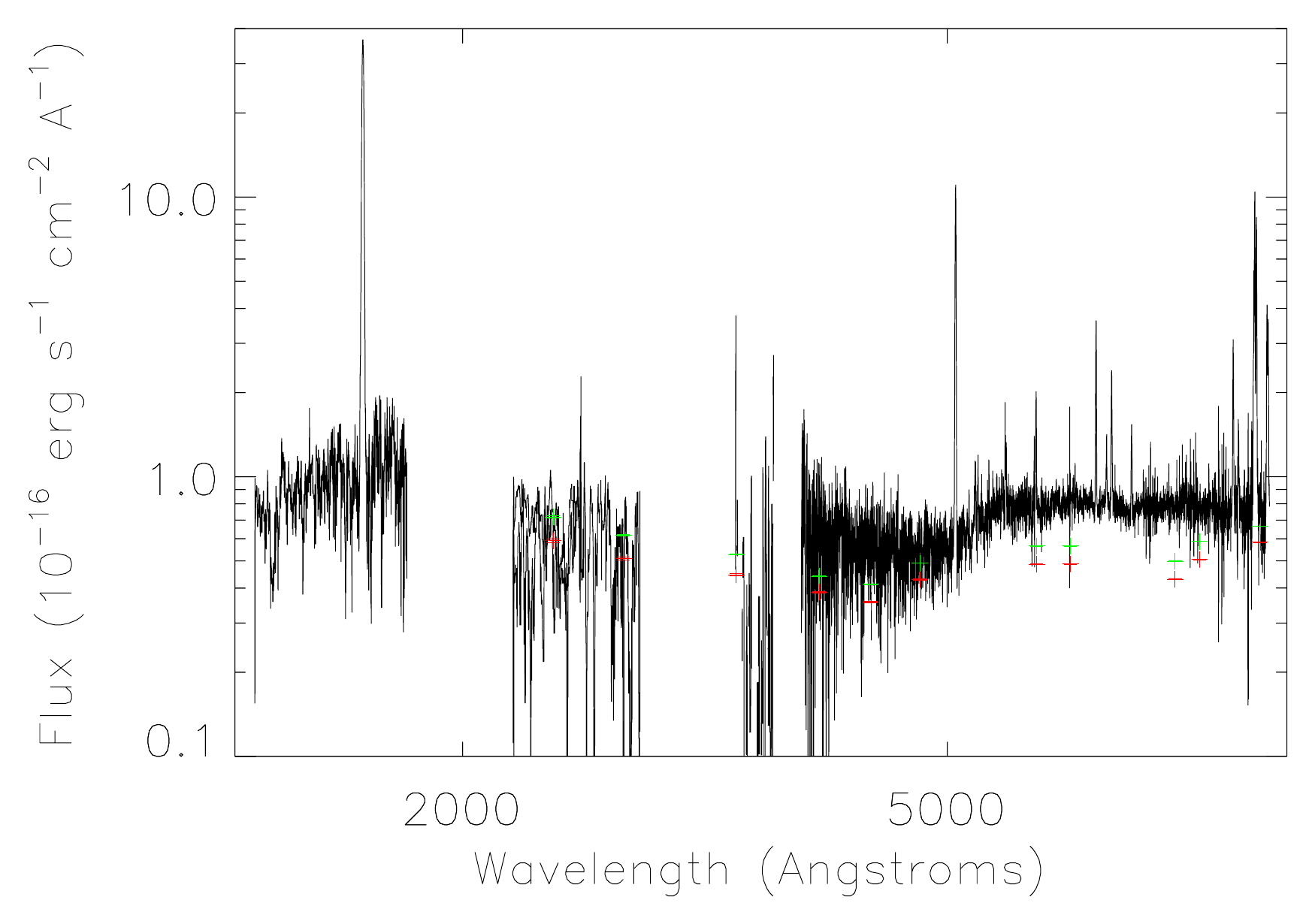}
\caption{\footnotesize
A composite spectrum of the central region of the RXJ1532.9+3021 BCG, from left to right, COS-FUV, COS-NUV, SDSS-optical. Overplotted in green and red points 
are the total broad-band flux estimates of the central $2\arcsec$ diameter aperture based on the CLASH HST photometry from WFC3 and ACS. The green points are the net fluxes, with only a sky background subtracted, as reported in Table~\ref{table:Phot}. The red points represent 
the fluxes with a background estimation made from a concentric $\Delta R =3\arcsec$ annulus (dominated by the BCG). Note that the SDSS/DR13 spectrum is based on a $3\arcsec$ diameter fiber, indicating the red light is less concentrated than the blue light, which is consistent with the appearance of the color image in Figure 1 as well.
  \label{figure:allspec}
  }
\end{figure*}

Note that small-aperture photometry of the individual knots yields NUV magnitudes 2 magnitudes fainter than that measured for the  large aperture, so most of the light in the large aperture  is coming from extended light, not any individual knot.
The 16-band CLASH  photometry for the two large apertures, centered on Knot A or Knot B, differs only at the 10\% level. 
This similarity shows that the small offset in aiming the final COS NUV observation makes very little difference. 
The net flux from the large-A aperture was converted from AB magnitudes to F$_\lambda$ units using the reported pivot wavelengths for each HST filter, and 
is plotted in Figure~\ref{figure:allspec} as individual points. This figure also includes the COS FUV and NUV spectra and 
the SDSS fiber spectrum, which represents a $3\arcsec$ diameter aperture, somewhat larger than the aperture used for the HST photometry or spectra. The correspondence with the HST broadband photometry for a similar region is quite good.

While investigating the cause of the aperture position difference between the June and December observations, 
we discovered that the raw fluxes based on photometry for the COS acquisition images varied from one observation to the next.  This variability led us to suspect 
intrinsic source variability (which would have been quite interesting if true), but subsequent investigation showed that vignetting by the COS acquisition camera is more likely to be the dominant culprit for the flux variations in the acquisition image. 
The vignetting correction is significant and too crude for our photometry on the finder image to improve over 
 the WFC photometry. But because of the vignetting uncertainty and the guide star difference we discussed earlier in this section, the last COS NUV observation was
centered on a position slightly offset ($\sim 0.4\arcsec $)
from the other COS pointings. The offset did not affect the fluxes in COS noticeably, because both clumps A and B are included in the high-throughput center of the COS aperture and in the resulting spectra for both pointings. So while the COS aperture shifted slightly in location ($0\farcs4$) from the FUV to the last NUV observation, we can treat it as an observation of nearly the same region because the same light sources are included in the COS aperture, 

Despite the photometric issues with the COS acquisition camera, its sampling of the HST Point Spread Function is
exquisite, at $0.0235\arcsec$ per pixel. We aligned and summed three acquisition images and show the result on the right-hand side of Figure~\ref{figure:COSFinder}, for comparison with the CLASH F275W image. Knot D is unresolved, with a FWHM of $0.07\arcsec$. Knot A is resolved, a single clump of about $0.7\arcsec$ (3.5 kpc) 
FWHM. Knot B, however, is clearly resolved into
a complex blend of sources, elongated approximately $0.5\arcsec$ (2.5 kpc) from the NE to the SW. Knot C remains compact with irregular structure at the edge of the acquisition field. The diffuse UV emission is lumpy, with some evidence of tails or filaments. Dark regions, which appear as dark grey or black in Figure~\ref{figure:COSFinder}
in both the COS acquisition image and the UV/CLASH data, lie 
just north of Knot A and to the SW of Knot B may correspond to dust features. 

\section{Discussion \label{discussion}}

The UV spectra of the central region of  RXJ1532.9+3021 are consistent with a continuous star formation rate of 
about 3-10 solar masses per year within a region with a diameter of $10$ kpc, uncorrected for intrinsic attenuation. 
Correcting for extinction implied by an $E(B-V)\sim0.2$ (consistent with the SDSS spectrum and with the Spectral Energy Distribution (SED) 
analysis by \citet{2015ApJ...813..117F} and comparison to the rates inferred from \halpha detected from a smaller region 
by our STIS observations and from a somewhat larger region by SDSS
would lead to star formation rates of up to 40-80 solar masses per year from the central 10 kpc.  Individual UV
star-forming knots are easily resolved by the COS acquisition image to be $0.3-0.7\arcsec$ or about $1.5-2.1$ kpc across.
Emission from at least the brightest two most central knots and extended emission contribute to the COS spectra.
The estimated star formation rate from the central 10 kpc is consistent with broadband CLASH photometry 
estimates of 100-150 solar masses per year over a larger, $\sim300$ kpc$^2$ sky area \citep{2015ApJ...813..117F}. 
The star formation is concentrated in, but not limited to, the central 10 kpc. 

The spectra from this region are similar to rest-frame UV spectra of Lyman Break Galaxies.  
We note that the rest-frame \lyalpha EQW seen here ($\sim 200\rm{\AA}$) would be among the highest \lyalpha EQWs seen
in the spectra of Lyman Break Galaxies \citep{2003ApJ...588...65S}.
 A couple of distinct stellar photospheric features from 
hot main-sequence stars are also visible in the UV continuum,  \ion{C}{3} 1174 \AA~ in particular. 
\lymanb is detected in absorption with a similar linewidth and redshift as the \lyalpha
emission line. These hydrogen lines are consistent with a wind $\sigma \sim 520\pm11$ km~s$^{-1}$, very
similar to winds seen around distant starburst galaxies \citep[e.g.,][]{2009ApJS..181..272G,2011ApJ...730....5H}.

High \lyalpha EQW is often associated with star-forming galaxies with systemically
lower metallicities \citep{2011ApJ...733..117F}. This \lyalpha system and other such systems  
residing in relatively high-metallicity BCGs might be counter-examples to this pattern. This galaxy is also
clearly not forming stars in a disk, given the morphology of the UV filaments and knots. 

We know that the presence of cold gas (as well as star formation and powerful radio sources) in BCGs is
strongly correlated with the thermodynamic state of the hot gas in the cluster 
core \citep[e.g.,][]{2008ApJ...683L.107C,2008ApJ...687..899R}. Some of the star-forming gas 
may therefore originate from low-entropy ICM gas dredged up from very center of the cluster, which then precipitates,
cools, and forms stars. The current {\em specific} star formation for these systems is quite low since the
stellar masses are already quite large; in fact, these galaxies have the most mass in stars or matter over
any other galaxies in the universe. 
However, stars produced by this mode of star formation -- which does not require a gas-rich merger or secular disk instabilities -- 
interestingly, would join the spheroidal stellar population of a galaxy. 
Therefore, BCGs such as this one are providing a  relatively nearby example of spheroidal star formation possibly being regulated by 
AGN feedback.  

 Cold gas may also be stripped from donor galaxies, but we regard this possibility as 
unlikely. Studies of BCGs with similar \halpha emission-line properties and star formation rates of order 100 M$_\odot$ yr$^{-1}$
also have considerable amounts of molecular hydrogen, 
10 billion M$_\odot$ or more \citep[e.g.,][]{2001MNRAS.328..762E,2011ApJ...738L..24V}, an amount also consistent
with the masses in other galaxies with similar star formation rates \citep{2011ApJ...738L..24V}. This amount
exceeds the amount of molecular gas in the Milky Way (from \citealt{2015ARA&A..53..583H}) by an order of magnitude. Given the lack of 
a massive donor suspect near the BCG and the similarity of this BCG compared to other BCG galaxies with direct molecular
gas detections, we are dubious that the cold gas fueling the star formation rates could have been stripped from a donor. 

Our limit on  \ion{O}{6} emission constrains the amount of cooling gas concurrently fueling star formation
in this system. Detections of \ion{O}{6} from knots in other similar systems such as the 
Phoenix Cluster at $z\sim0.6$  \citep{2015ApJ...811..111M} and
Abell 1795 at $z \sim 0.05$ \citep{2014ApJ...791L..30M} 
have shown the radiative cooling rates based on \ion{O}{6} emission-line detections can 
exceed the local star formation rate. In those works it was suggested that the star formation
is therefore inefficient. In an off-center knot in M87, the detection of 100,000K gas via \ion{C}{4} emission lines, 
along with the lack of recent star formation, led
to the conclusion that conduction may be an important process in sustaining an intermediate-temperature 
interface between the X-ray gas and the  cooler gas, since neither recently-formed stars nor AGN were present \citep{2009ApJ...704L..20S,2012ApJ...750L...5S}.

But the observations we report here of a very similar system have rather different implications.  We see strong signatures of
star formation and relatively little evidence for 100,000~K gas. 
Hotter $10^7$~K gas is certainly present, along with significant cavities lying a little farther from the center, 
outside the emission line filaments, as indicated by Chandra
X-ray observations \citep{2013ApJ...777..163H}.  This brighest cluster galaxy and its molecular, ionized, and X-ray
gas content and distributions are basically identical to other brightest cluster galaxies with radio sources, excess UV and FIR light, and 
extended, filamenatary optical emission line gas. Consequently, we conclude that RXJ1532.9+3021 is in a different phase
of an episodic feedback process.

In recent simulations of idealized AGN feedback by \citet{2015ApJ...811...73L}, the rate of gas cooling in the simulation does indeed drop
below the star formation rate in the latter part of the bursts. It is too soon to conclude that we are witnessing the closing phase of
an episode of AGN outburst and the star formation associated with that event, but it is interesting to think about how to use an ensemble of similar measurements for
other clusters to test such simulations, as well as how to use simulations to interpret individual galaxy measurements.

We suggest that the most plausible interpretation is that the star formation rate in a given BCG region 
does not track the current radiative cooling rate in lockstep, even in a system in which the ICM fuels the star formation.
This interpretation is not completely speculative,  since simulations of AGN feedback triggered by condensation suggest
that the condensation events are episodic 
\citep{2012ApJ...746...94G,2015ApJ...811..108P,2015ApJ...811...73L,2016arXiv160303674M,2016ApJ...829...90Y}.
Such an interpretation would suggest that
RXJ1532.9+3021 is perhaps a system at the end of a major outburst, while other \ion{O}{6}-detections may be
revealing systems in the earlier phases. The details from any given simulation undoubtedly depend on exactly
how the simulators implement star formation, AGN feedback, and cooling in their codes, but
one would expect scatter in the ratios of \ion{O}{6} to SFRs for systems where the source of kinetic energy
varies in power and direction. Simulators  
do not yet have statistical predictions for time-averaged scatter in those observables since simulations with resolution sufficient to follow
the cold gas are typically of only single systems. Future UV observations of emission-line
regions in BCGs would provide very useful constraints for probing the timescales of AGN activity, star formation, and 
mass condensation. Ideally, these observations should include regions without optical AGN, since disentangling the effects
of an optically-visible AGN from that of cooling gas would be challenging.

\section{Conclusions \label{conclusions}}

The COS FUV spectrum of the core of the central galaxy in RXJ1532.9+3021
shows a strong \lyalpha line (rest EQW $\sim200$\AA), stellar UV continuum, and no other emission lines (such as \ion{O}{6}, \ion{N}{5}, or \ion{C}{4}).
The stellar continuum of the FUV spectrum shows \lymanb  absorption (with a $N_{HI} \sim  4.9 \times 10^{15}$ cm$^{-2}$), 
likely due to stellar winds,  and stellar absorption features
characteristic of the photospheres of hot stars. The \lyalpha emission line is likely to have additional broadening because it is
being scattered by hydrogen atoms in a wind. A comparison of the FUV spectrum with a composite spectrum of $z=3$ 
Lyman break galaxies corroborates the
identification of these features with recently formed stars. We present a long-slit STIS spectrum showing \halpha and
relatively weakly-detected [\ion{N}{2}], which is also consistent with gas photoionized by hot stars. The width of the \halpha
feature is  $\sigma \sim 220$ km s$^{-1}$,  less broad than the \lyalpha feature ($\sigma \sim 500$ km s$^{-1}$) seen from a similar region
of the galaxy. The most likely explanation is that the \lyalpha photons are far more strongly scattered by the
winds in the region. The ratio of \halpha to \lyalpha is consistent with a moderate level of dust
extinction. The EQW of \lyalpha is extraordinary, especially since this BCG is neither low-metallicity nor high-redshift.

All of these properties are consistent with active 
star formation. The inferred UV star formation rate, uncorrected for any obscuration, is $\sim 3-10$ solar masses per 
year in the central kiloparsec, while the rate much less affected by extinction from H$\alpha$ is 20-80 solar masses per year, depending
on the aperture. The overall rate for the entire galaxy, as estimated from previous Spitzer 
MIR photometry, is $\sim 100$ solar masses per year, a value consistent with that inferred from 
an analysis of the 16 bandpass CLASH photometry SED \citep{2015ApJ...813..117F}.

Our analysis of the COS spectra places a stringent upper limit on the amount of concurrently cooling gas in the
system, based on the absence of \ion{O}{6} emission lines (and to some degree, the absence of a \ion{C}{4} emission line).
In the RXJ1532.9+3021 system, the gas mass cooling rate inferred from \ion{O}{6} upper limits is significantly less
than the star formation rate (extinction-corrected UV, \halpha, FIR, and SED).  This result was unexpected, since 
previous rest-UV spectrocopic observations of similar emission-line BCGs in X-ray
cool core clusters displayed much more  luminous \ion{O}{6} and \ion{C}{4} 
emission lines than expected if the gas cooling rates were equal to the star formation rates. 
Our interpretation is that this BCG is indeed a similar system to the Phoenix
Cluster, M87, or Abell 1795, with similar mechanisms regulating AGN feedback, radiative cooling, and star formation, 
but in a phase in which the star formation rate exceeds the gas cooling rate. All of those
systems are forming stars without the assistance of a gas-rich merger or the presence of a detectable gaseous
disk. Such systems may be examples of how some stars can enter a galaxy's spheroidal population without ever belonging
to a disk. Similar UV spectroscopic observations of a larger sample
of emission-line BCGs may reveal an ensemble of states, in which accretion (or radiative cooling) does not precisely
track the star formation rates. The proportions of galaxies in each state for a well-chosen sample will give us
better observational constraints on the timescales of feedback, cooling, and heating in these systems, which in turn
will lead to better understanding of the mechanisms underlying AGN feedback.

\acknowledgments

%% See the AASTeX Web site at http://www.journals.uchicagoEdu/AAS/AASTeX

MD and TC were partially supported by an STScI/NASA award HST-GO-13367. MD, TC, and MP were
partially supported by an STScI/NASA award HST-GO-12065.
This research made use of ADS and NED.
This research has made use of the VizieR catalogue access tool, CDS, Strasbourg, France. 
The original description of the VizieR service was published in A\&AS 143, 23.
The first author is grateful the useful advice of Dr. Steve Penton of STScI in planning 
the COS observations and providing a second verification of the COS acquisition data.

Funding for the SDSS and SDSS-II has been provided by the Alfred P. Sloan Foundation, the Participating Institutions, the National Science Foundation, the U.S. Department of Energy, the National Aeronautics and Space Administration, the Japanese Monbukagakusho, the Max Planck Society, and the Higher Education Funding Council for England. The SDSS Web Site is \url{http://www.sdss.org/}. The SDSS is managed by the Astrophysical Research Consortium for the Participating Institutions. The Participating Institutions are the American Museum of Natural History, Astrophysical Institute Potsdam, University of Basel, University of Cambridge, Case Western Reserve University, University of Chicago, Drexel University, Fermilab, the Institute for Advanced Study, the Japan Participation Group, Johns Hopkins University, the Joint Institute for Nuclear Astrophysics, the Kavli Institute for Particle Astrophysics and Cosmology, the Korean Scientist Group, the Chinese Academy of Sciences (LAMOST), Los Alamos National Laboratory, the Max-Planck-Institute for Astronomy (MPIA), the Max-Planck-Institute for Astrophysics (MPA), New Mexico State University, Ohio State University, University of Pittsburgh, University of Portsmouth, Princeton University, the United States Naval Observatory, and the University of Washington.

%% in parentheses, after the keyword, but they will not be verified.

{\it Facilities:} \facility{HST (ACS, WFC3, COS, STIS), Sloan}

\vspace{1em}
\bibliography{CLASH_UV}

\clearpage

\end{document}